\documentclass[aps,prb,twocolumn,superscriptaddress,longbibliography]{revtex4-2}

\usepackage[colorlinks=true,citecolor=blue]{hyperref}
\usepackage{graphicx}
\graphicspath{ {./Figures/} }
\usepackage{amsmath}
\usepackage{amssymb}
\usepackage{bbold}
\usepackage{enumerate}
\usepackage[dvipsnames]{xcolor}
\usepackage{physics}

\usepackage{orcidlink}

\newcommand{\co}[2]{#2}
\begin{document}

\author{Aleksander Sanjuan Ciepielewski\orcidlink{https://orcid.org/0000-0001-5883-4903}}
\affiliation{International Research Centre MagTop, Institute of Physics, Polish Academy of Sciences, Al. Lotnik\'ow 32/46, 02-668 Warsaw, Poland}

\author{Jakub Tworzyd\l{}o\orcidlink{https://orcid.org/0000-0003-3410-5460}}
\affiliation{Faculty of Physics, University of Warsaw, ulica Pasteura 5, 02-093 Warsaw, Poland}

\author{Timo Hyart\orcidlink{https://orcid.org/0000-0003-2587-9755}}
\affiliation{International Research Centre MagTop, Institute of Physics, Polish Academy of Sciences, Al. Lotnik\'ow 32/46, 02-668 Warsaw, Poland}
\affiliation{Department of Applied Physics, Aalto University, 00076 Aalto, Espoo, Finland}
\affiliation{Computational Physics Laboratory, Physics Unit, Faculty of Engineering and Natural Sciences, Tampere University, FI-33014 Tampere, Finland}

\author{Alexander Lau\orcidlink{https://orcid.org/0000-0001-6671-8056}}
\affiliation{International Research Centre MagTop, Institute of Physics, Polish Academy of Sciences, Al. Lotnik\'ow 32/46, 02-668 Warsaw, Poland}

\title{Transport signatures of van Hove singularities in mesoscopic twisted bilayer graphene}

\date{\today}

%-------------------------------------------------------

\begin{abstract}
	Magic-angle twisted bilayer graphene exhibits quasi-flat low-energy bands with van Hove singularities close to the Fermi level. 
	These singularities play an important role in the exotic phenomena observed in this material, such as superconductivity and magnetism, by amplifying electronic correlation effects. 
	In this work, we study the correspondence of four-terminal conductance and the Fermi surface topology as a function of the twist angle, pressure, and energy in mesoscopic, ballistic samples of small-angle twisted bilayer graphene. We establish a correspondence between features in the wide-junction conductance and the presence of van Hove singularities in the density of states.
	Moreover, we identify additional transport features, such as a large, pressure-tunable minimal conductance,  conductance peaks coinciding with non-singular band crossings, and unusually large conductance oscillations as a function of the system size. 
	Our results suggest that twisted bilayer graphene close the magic angle is a unique system featuring simultaneously large conductance due to the quasi-flat bands, strong quantum nonlinearity due to the van Hove singularities and high sensitivity to external parameters, which could be utilized in high-frequency device applications and sensitive detectors.
\end{abstract}

\maketitle

\section{Introduction}

\co{TBLG: Why is it an interesting subject of study? Correlated physics due to flat bands around the magic angle and  topological phases. Selection of important experimental findings. Introduce van Hove singularities.}

Twisted bilayer graphene (TBLG) has attracted a lot of attention because of its fascinating phenomena at certain twist angles~\cite{morell2010, Bistritzer2011a, Kim2017, Cao2018a, Cao2018b, Liu2019, Yuan2019, Song2019, Lu2019, Tarnopolsky2019}, commonly referred to as \emph{magic} angles.
At these twist angles, the material exhibits exceptionally flat energy bands, leading to an enhancement of electronic interactions, and appearance of superconductivity and other correlated phases ~\cite{Ojajarvi2018, Cao2018a, Cao2018b, Liang2018, Wu2018, PeltonenPhysRevB.98.220504, Liang2019, Yankowitz2019,  Lu2019, Hazra2019, Hu2019, Fang2020, Julku2020}.
Besides their flatness, the low-energy bands also feature van Hove singularities (VHSs) in the density of states (DOS)~\cite{vanHove1953} close to the Fermi level. At ordinary VHSs, the DOS diverges logarithmically, whereas it diverges with a power law in higher-order VHSs ~\cite{Yuan2019}. 
The importance of VHSs for the correlated phenomena observed in TBLG has been pointed out theoretically~\cite{Wu2018, Liang2018, Liang2019} and their existence has been confirmed in scanning tunnelling spectroscopy experiments~\cite{Li2010, Kerelsky2019, Choi2019}.
 
\co{Previous studies similar to ours.  We fill the gap by considering multi terminal transport in mesoscopic TBLG samples. Draw connection between transport signatures and Fermi surface topology around the magic angle.}

Most of the experimental and theoretical research on TBLG has so far focused on observables in the thermodynamic limit and on transport in macroscopic samples in the semiclassical regime  \cite{WudasSarma19, HwangdasSarma}. The quantum transport studies have so far addressed specific questions such as the angle-dependent minimal conductivity, disorder effects, and emergent magnetic textures in driven TBLG~\cite{Andelkovic2018, Padhi20, Bahamon2020}, as well as transport in crossed graphene nanoribbons, where the scattering region is smaller than the magic-angle moir{\'e} unit cell~\cite{Zhou2010, Brandimarte2017, Sanz2020}. However, it remains an outstanding challenge to understand the effects of quasi-flat bands, VHSs and the rich variety of possible Fermi surface topologies on the quantum transport in mesoscopic TBLG samples, where the transport characteristics can be probed in an energy-resolved fashion. Recently, first steps towards this direction have been taken by identifying quantum transport signatures of VHSs in TBLG in the regime of intermediate twist angles $3^\circ \lesssim \theta \lesssim 10^\circ$ assuming semimetallic leads~\cite{Olyaei2020}. In the semimetallic leads the DOS goes to zero at the quasi-flat band energies thereby hindering the identification of quantum transport signatures related to the VHSs within the quasi-flat bands close to the magic angle $\theta \sim 1^\circ$. 
Here, we go further into this direction by studying four-terminal conductance in mesoscopic, ballistic TBLG samples, containing approximately one million sites in the scattering region, around the first magic angle. Importantly, we obtain a higher energy resolution by using metallic leads and, therefore, we are able to study the effects of the VHSs and of the Fermi surface topology of the quasi-flat bands on the quantum transport.

We find that the low-energy quantum transport in TBLG close to the magic angle is affected by several factors.
We demonstrate that by tuning the twist angle or pressure to flatten the energy bands the system can support considerably larger minimal conductance than monolayer and Bernal-stacked bilayer graphene devices~\cite{Katsnelson2006, Tworzydo2006, Katsnelson2006bilayer, Snyman2007, note_bilayer_conductance}. 
We further link energy-dependent conductance signatures to different VHSs in the bulk DOS and to non-singular band crossings, and we observe unusually large conductance oscillations as a function of the system size.
Our findings put forward that TBLG close the magic angle is an exceptional system combining  large conductance  originating from quasi-flat bands with strong quantum nonlinearity from VHSs and high sensitivity to external parameters.
We propose that these properties could be utilized in compact high-frequency devices and sensitive detectors. Such applications  could utilize the quantum twisting microscope technology offering the possibility to locally and continuously tune pressure and twist angle in TBLG devices \cite{Ilani22}.

\co{Structure of the paper.}

The structure of the paper is as follows.
In Sec.~\ref{sec:model-and-setup} we introduce the four-terminal geometry for the transport calculations and the TBLG model used in this work.
We continue by presenting in Sec.~\ref{sec:zero-energy-conductance} the minimal conductance as a function of the twist angle, pressure, and the system size.
In Sec.~\ref{sec:conductance-and-VHS} we identify the effects of VHSs and the Fermi surface topology on the energy dependence of the conductance. Finally, we discuss possible high-frequency device applications in Sec.~\ref{device-applications} and we summarize our results in Sec.~\ref{sec:conclusion}.

\section{Model and setup for transport calculations}
\label{sec:model-and-setup}

\subsection{Twisted bilayer graphene}

\co{Review the geometry of untwisted and twisted bilayer graphene.}

There are two common configurations for untwisted bilayer graphene (BLG)~\cite{CastroNeto2009, Rozhkov2016}:
In AA-stacked BLG, corresponding atoms from different layers are on top of each other.
In Bernal AB-stacked BLG, on the other hand, one of the layers is shifted relative to the other, such that some atoms from one layer lie at the centers of the hexagons formed by the other layer and vice versa.
Both configurations have in common that their primitive lattice vectors are identical to those of single-layer graphene.
In TBLG, starting from one of these configurations, the two layers are rotated relative to each other by an angle $\theta$ around a fixed point in space~\cite{Andrei2020}.
As a consequence, a moir{\'e} pattern emerges, which breaks the translational symmetry of the individual graphene layers.
For certain angles, however, the two layers form a periodic moir\'{e} honeycomb superlattice, whose primitive lattice vectors and lattice constant are angle-dependent.
These \emph{commensurate} twist angles have the following form~\cite{dosSantos2012}
\begin{equation}
    \cos{\theta} = \frac{3m^2 + 3mr + r^2/2}{3m^2 + 3mr + r^2},
\end{equation}
where $m$ and $r$ are coprime positive integers.
Based on this notation, the moir{\'e} lattice constant is~\cite{Zou2018}
\begin{equation}
    a = \frac{a_0}{2\sin{\frac{\theta}{2}}} \frac{r}{\sqrt{\mathrm{gcd}(r,3)}},
\end{equation}
with the lattice constant $a_0=2.46\,\textrm{\AA}$ of single-layer graphene and $\mathrm{gcd}(p,q)$ denotes the greatest common divisors of the integers $p$ and $q$.
Generally, the smaller the twist angle, the larger the moir{\'e} lattice constant.
At the first magic-angle $\theta = 1.05^\circ$ ($m = 31$ and $r = 1$), for instance, we have $ a\approx  15\,\textrm{nm}$.

\subsection{Four-terminal transport setup}

\begin{figure}[t]
\includegraphics[width=0.485\textwidth]{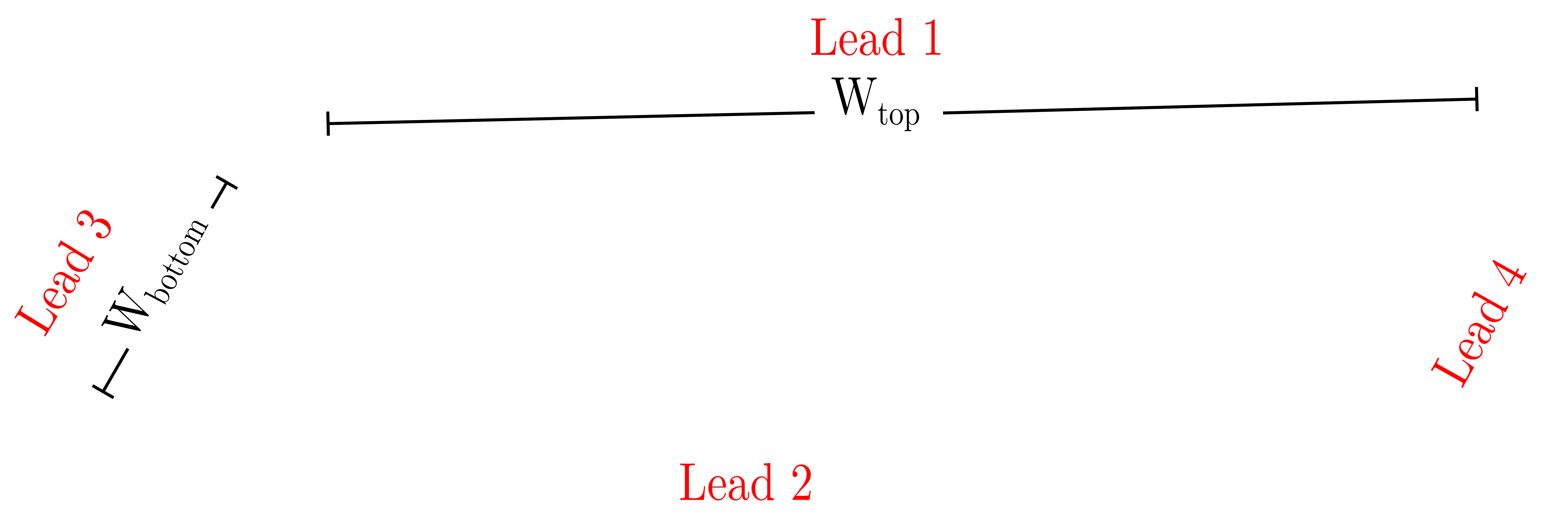}
\vspace{-0.8cm}
\caption{Twisted bilayer setup: two crossed graphene nanoribbons with leads (red) form a bilayer region.
The top layer is twisted by an angle $\theta$ relative to the bottom layer around the center of the overlap region.
We show a system at the magic angle $\theta=1.05^\circ$, and widths of $W_\mathrm{top}= 250\,\mathrm{nm}$ for the top ribbon and $W_\mathrm{bottom}=50\,\mathrm{nm}$ for the bottom ribbon, corresponding to the largest system considered in this work.}
\label{fig:setup}
\end{figure}

\co{Describe setup of two crossed graphene ribbons and how the leads are attached.}

In this study, we consider ballistic quantum transport through a TBLG region formed by two crossed graphene ribbons, as illustrated in Fig.~\ref{fig:setup}.
Such a setup could be realized in the lab with state-of-the-art experimental techniques~\cite{Jiao2010} and the recent advances allow also to tune the twist angle and pressure locally and continuously~\cite{Ilani22}.
We choose the ribbons to both have armchair terminations to avoid contributions from edge states to the electronic transport, as we aim to study bulk signatures in this work.
This results in a parallelogram-shaped overlap region between the two ribbons. 
In our setup, the top ribbon can be twisted by an angle $\theta$ around the center of the bilayer region.
The ribbons are placed such that for $\theta=0$ the two layers are stacked in an AA fashion in the overlap region.
We note that we obtain qualitatively similar results as shown in this paper if we start from an AB-stacked overlap region instead (not shown).
Moreover, the continuation of the ribbons outside the bilayer region defines four semi-infinite, monolayer graphene leads, which we use for our transport calculations (see Fig.~\ref{fig:setup}).
In contrast to previous work~\cite{Olyaei2020}, we use metallic leads by tuning the chemical potential outside the scattering region far away from the Dirac-point energy (see below).

\subsection{Tight-binding model}

\co{Brief motivation for our modelling approach}

In the literature, various modelling approaches have been used to study the low-energy properties of TBLG, such as continuum models~\cite{dosSantos2007,Mele2010,Bistritzer2011a}, ab-initio calculations~\cite{Sanz2020, Zhou2010, Brandimarte2017}, or tight-binding models~\cite{Cao2021, morell2010, Lin2018, Moon2012, Po2019}.
Here, we aim to investigate multi-terminal electronic transport in mesoscopic samples with a focus on the small-angle regime. 
Continuum models are long-wavelength low-energy theories and, therefore, cannot fully capture the details at length and energy scales relevant to transport in mesoscopic systems. 
In the case of ab-initio calculations, it is computationally expensive to model samples sufficiently large to overcome finite-size effects at small twist angles, and thereby the effects attributable to the electronic properties of the bulk are necessarily obscured.
Tight-binding models, on the other hand, are able to accurately capture the low-energy electronic properties of TBLG over a wide range of twist angles~\cite{Lin2018, Moon2012} while having the advantage of being computationally cheaper to scale up.
For this reason, we adopt a tight-binding model approach in this work.

\co{General form of the Hamiltonian and single-layer part}

The generic form of a Hamiltonian for a weakly-coupled bilayer system, such as TBLG, is $H = H_1 + H_2 + H_{12}$, where $H_{1,2}$ are the Hamiltonians of the individual layers and $H_{12}$ contains interlayer-coupling terms~\cite{Moon2012, Olyaei2020, Lin2018}.
For the individual layers and the leads, we use the common nearest-neighbor tight-binding Hamiltonian of graphene~\cite{CastroNeto2009}
\begin{eqnarray}
    H_m &=& -t\sum_{\langle i,j\rangle,\sigma} c_{im\sigma}^\dagger c_{jm\sigma} - \mu \sum_{i,\sigma} c_{im\sigma}^\dagger c_{im\sigma},
\end{eqnarray}
where $c_{im\sigma}^\dagger$ ($c_{im\sigma}$) creates (annihilates) a $p_z$ electron with spin $\sigma=\uparrow,\downarrow$ at lattice site $\mathbf{r}_i$ of the $m$-th layer, $t=3.09\,\mathrm{eV}$ is the nearest-neighbor hopping amplitude~\cite{Lin2018}, and $\mu$ is the chemical potential.
We set $\mu=-2\,\mathrm{eV}$ in the leads, such that the leads are metallic, and $\mu=0$ throughout the scattering region.

For the interlayer part of the Hamiltonian, we follow Ref.~\onlinecite{Lin2018} by using
\begin{equation}
    H_{12} = -\sum_{\langle i,j\rangle,\sigma} t'(r_{ij})\,c_{i,2,\sigma}^\dagger c_{j,1,\sigma} + \mathrm{H.c.},
\end{equation}
where $r_{ij}=|\mathbf{r}_i-\mathbf{r}_j|$ is the in-plane distance between two lattice sites in different layers at positions $\mathbf{r}_i$ and $\mathbf{r}_j$, respectively, and $t'(r)$ is the isotropic interlayer hopping integral given by
\begin{equation}
    t'(r) = V_{pp\sigma}^0\, e^{-\left(\sqrt{r^2+d_0^2}-d_0\right)/\lambda} \frac{d_0^2}{r^2+d_0^2},
    \label{eq:interlayer_coupling}
\end{equation}
with the nearest-neighbor interlayer coupling $V^0_{pp\sigma} = 0.39\,\mathrm{eV}$, the distance between the graphene layers $d_0=3.35\,\textrm{\AA}$, and the decay parameter $\lambda$~\cite{Lin2018}.
The form of the interlayer hopping term is based on a Slater-Koster approximation for the overlap integrals between the $p_z$ orbitals in different layers~\cite{SlaterKoster1954}.
It is generally found that longer-range interlayer hopping terms have to be taken into account to capture the electronic bands of TBLG in a wide range of angles~\cite{Lin2018}. 
In accordance with Ref.~\onlinecite{Lin2018}, we use $\lambda=0.27\,\textrm{\AA}$, which reproduces the well-known band structures of AA- and AB-stacked BLG.
Making use of the rapidly decaying nature of the hopping integral $t(r)$ in Eq.~\eqref{eq:interlayer_coupling}, we further neglect interlayer terms with $r > 5\,\textrm{\AA}$, which is sufficient to accurately capture the bands of TBLG at the first-magic angle.

\co{We neglect interaction effects.}

We note that we neglect interaction effects in our model.
These are generally important for the ground-state properties of TBLG when the Fermi level is within the quasi-flat bands and, thus, correlated phases emerge.
However, this would lead to a reconstruction of the energy bands thereby obscuring the origin of the enhanced interactions, such as VHSs.
We therefore restrict our study to a non-interacting description of the system, which is a good approximation as long as the Fermi level is tuned away from the quasi-flat bands. The transport can still be studied in an energy-resolved manner by tuning the voltage bias. In this case, only nonequilibrium quasiparticles are occupying the flat-band states so that interaction effects are not expected to be as important as in the case of equilibrium flat-band systems.
Alternatively, it is also possible to screen the interactions so that a non-interacting description becomes more accurate \cite{Veyrat2020, Stepanov2020}.
Furthermore, the correlated phases appear only at low temperatures. 

\co{Use of kwant and numerical details for the finite system (e.g. system sizes)}
For the four-terminal transport calculations of the device shown in Fig.~\ref{fig:setup} we use the quantum transport Python package kwant~\cite{Groth2014}.
To be able to capture bulk effects in the transport calculations, the size of the bilayer region needs to be at least on the order of several moir\'{e} unit cells.
Using an interlayer-hopping cut-off as explained above enables us to efficiently study systems with close to $10^6$ lattice sites.
In particular, for Fig.~\ref{fig:conductance_angle_and_interlayer_dependance} we perform calculations for samples of $3.2\times 16$ magic-angle moir\'{e} unit cells, which is equivalent to $40\,\mathrm{nm} \times 200\,\mathrm{nm}$. For Fig.~\ref{fig:conductance_dos_spectrum} we increased the size of the samples to $4\times 20$ magic-angle moir\'{e} unit cells ($50\,\mathrm{nm} \times 250\,\mathrm{nm}$).
For such samples, the bottom-layer nanoribbon represents a long and narrow junction, whereas the top-layer nanoribbon realizes a short and wide junction (see Fig.~\ref{fig:setup}).
Our setup also allows us to study the effect of pressure which changes the value of the interlayer coupling.

\co{Numerical details for spectrum calculations and DOS}
We further aim to draw a connection between transport signatures and spectral features of the bulk, in particular VHSs.
Therefore, for commensurate twist angles, we impose periodic boundary conditions on the moir\'{e} unit cell and calculate 
the band structure.
This enables us to extract the bulk DOS and the Fermi surface of the system at fixed energies, which we use to pin down the VHSs.
We note, however, that the observed transport signatures do not depend on whether the system is commensurate or incommensurate.
Throughout this work, we align $E=0$ with the energy of the Dirac points at the $\mathrm{K}$ and $\mathrm{K}'$ points of the moir{\'e} Brillouin zone (BZ).
For incommensurate twist angles, we interpolate linearly between the Dirac-point energies corresponding to the closest commensurate twist angles.

\section{Conductance at the Dirac-point energy}
\label{sec:zero-energy-conductance}

\co{Define Conductance tensor, discuss which components of $G_{ij}$ are independent.}

For a multi-terminal setup, the differential conductance $G$ is a tensor defined through $G_{ij} =dI_i/dV_j$, where $I_i$ is the current at the $i$-th lead and $V_j$ is the voltage at the $j$-th lead of the system.
Here, we consider a system with four leads, so our conductance tensor is represented by a $4\times 4$ matrix.
Moreover, our setup has an approximate $C_2$ symmetry, which is only weakly broken along the boundaries of the system due to its terminations.
Hence, the conductance tensor $G_{ij}$ has only four independent components, which we choose to be the intralayer conductances $G_{12}$ (short and wide junction) and $G_{34}$ (long and narrow  junction), as well as the interlayer conductances $G_{13}$ and $G_{14}$.

\begin{figure}[t]
\includegraphics[width=0.45\textwidth]{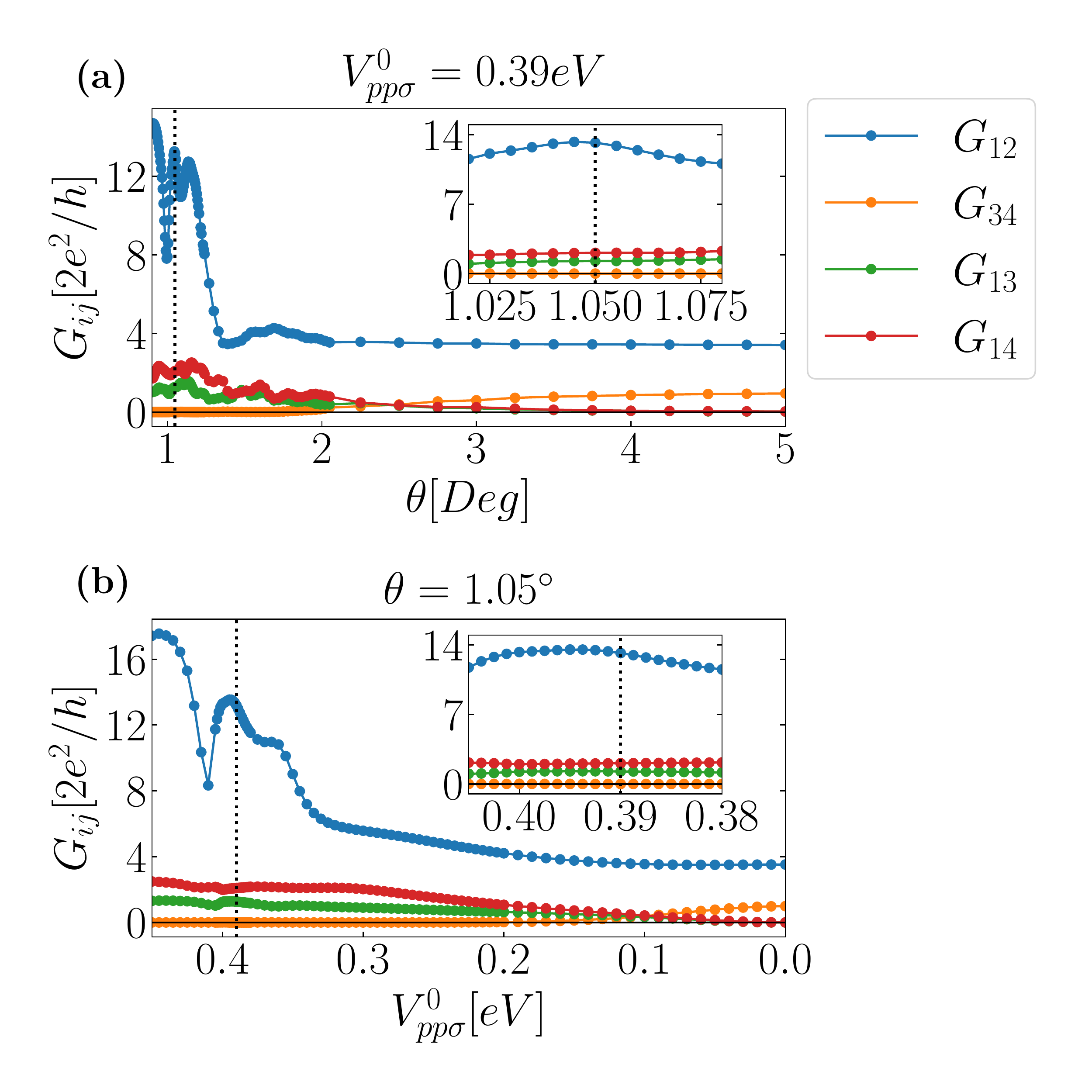}
\caption{Four-terminal minimal conductance $G_{ij}$ at $E=0$: (a) as a function of the twist angle $\theta$ at fixed interlayer coupling $V_{pp\sigma}^0=0.39\,\mathrm{eV}$, (b) as a function of the interlayer coupling $V_{pp\sigma}^0$ at the first magic angle $\theta=1.05^\circ$. 
We show the following components of the conductance matrix: $G_{12}$ (blue), $G_{34}$ (orange), $G_{13}$ (green), and $G_{14}$ (red).
The insets show the conductance close to (a) the magic angle $\theta=1.05^\circ$ and to (b) the equilibrium interlayer coupling $V_{pp\sigma}^0=0.39\,\mathrm{eV}$, which are indicated by the respective dotted black lines.}
\label{fig:conductance_angle_and_interlayer_dependance}
\end{figure}

\co{Discuss large and intermediate angles, single-layer limits, minimal conductance}
First, we compute the conductance $G_{ij}$ as a function of the twist angle $\theta$ at the  Dirac-point energy $E=0$, which we dub \emph{minimal conductance} of the quasi-flat band system in analogy with monolayer graphene.
In fact, for twist angles $\theta > 1.05^\circ$ the wide junction conductance typically has a local minimum at this energy. We note, however, that at smaller twist angles the conductance is no longer necessarily minimal at $E=0$ for reasons explained in Sec.~\ref{sec:conductance-and-VHS}. 
We show our results in Fig.~\ref{fig:conductance_angle_and_interlayer_dependance}(a).
At large angles, $\theta\gtrsim 4^\circ$, the conductance shows a universal behavior independent of the twist angle.
The short-junction conductance $G_{12}$ approaches a value corresponding to the minimal Dirac-point conductivity of single-layer graphene, namely $G_{12}\,W_\mathrm{bottom}/W_\mathrm{top}\approx 4e^2/\pi h$, 
which is in agreement with previous results in the literature~\cite{Tworzydo2006, Katsnelson2006, Andelkovic2018}.
The long-junction conductance $G_{34}$, on the other hand, approaches the quantized value of $2e^2/h$. 
We find that it corresponds to a single spin-degenerate, propagating bulk mode confined to the bottom layer, whose presence is attributed to the particular width of the nanoribbon~\cite{Nakada1996}. 
At the same time, the interlayer conductances $G_{13}$ and $G_{14}$ vanish.
Hence, the two nanoribbons are effectively decoupled and the conductance tensor decomposes into two independent two-terminal conductances. 
This is in agreement with previous results~\cite{Andelkovic2018}.

\co{Discuss small-angle regime and how the layers couple, highlight the short-junction conductance}

For small angles, $\theta\lesssim 4^\circ$, the interlayer conductances become nonzero clearly indicating coupling between the two nanoribbons.
This is also reflected in the behavior of the long-junction conductance $G_{34}$, which is suppressed, because the propagating bulk mode in the bottom layer can now interfere with the corresponding mode in the top layer thereby lowering the conductance. Close to the magic angle, $G_{34}$ becomes nearly zero.
On the contrary, the short-junction conductance $G_{12}$ is strongly enhanced as we get closer to the magic angle and deviates considerably from the single-layer value.
The behavior of the different conductance channels can be attributed to the formation of quasi-flat energy bands in small-angle TBLG:
the presence of the flat band enhances the DOS around $E=0$, while the Fermi velocities of the corresponding bulk modes are decreased.
A measurable enhancement of conductance requires a sufficiently large number of lead modes, which is why the wide-junction conductance $G_{12}$ shows the largest effect.
Therefore,  $G_{12}$ is a suitable quantity to probe 
how the van Hove singularities affect the transport properties through their enhancement of the DOS.
We note that there are also other factors influencing the conductance beyond the bulk DOS, as we will discuss below.

\co{Describe the behavior of the short-junction conductance at small angles}
%The inset of Fig.~\ref{fig:conductance_angle_and_interlayer_dependance}(a) shows the details of the conductance close to the magic angle.
The wide-junction conductance $G_{12}$ is generally enhanced towards smaller angles, but shows a non-monotonic behavior.
It features two pronounced maxima, one at $\theta=1.14^\circ$ and one at the magic angle $\theta=1.05^\circ$, and increases again for $\theta\lesssim1.0^\circ$.
We discuss this behavior in more detail below in the context of the correspondence between conductance and DOS.

\co{Discuss briefly the conductance as a function of interlayer coupling}
In Fig.~\ref{fig:conductance_angle_and_interlayer_dependance}(b), we show the minimal conductance also as a function of the nearest-neighbor interlayer coupling $V^0_{pp\sigma}$ at the magic angle $\theta=1.05^\circ$.
This parameter can be controlled by applying pressure~\cite{Yankowitz2019,Padhi2019}.
Overall, all components of the conductance tensor $G_{ij}$ exhibit the same qualitative behavior as in the case of a variation of the twist angle.
In particular, the conductances for vanishing interlayer coupling $V^0_{pp\sigma}\approx 0$ are the same as in the large twist-angle regime confirming the picture of effectively decoupled nanoribbons.
In the large-coupling regime, we further observe the same enhancement of the interlayer conductances and of the wide-junction conductance, while the long-junction conductance is suppressed.
Close to the magic angle, the similarities between the variations of these two parameters, $\theta$ and $V^0_{pp\sigma}$, even show a good quantitative agreement, as %we indicate in the inset of Fig.~\ref{fig:conductance_angle_and_interlayer_dependance}(b).
can be seen by comparing the insets of Figs.~\ref{fig:conductance_angle_and_interlayer_dependance}(a) and~\ref{fig:conductance_angle_and_interlayer_dependance}(b).
More generally, we find that also the corresponding bulk energy bands evolve in a similar way.
Our results suggest that a TBLG sample at an \emph{incommensurate} angle can be approximated by a sample at the closest \emph{commensurate} angle in combination with applied pressure.
Below, we will use this insight to continuously trace the evolution of VHSs as a function of the twist angle.

\begin{figure}[t]
\includegraphics[width=0.40\textwidth]{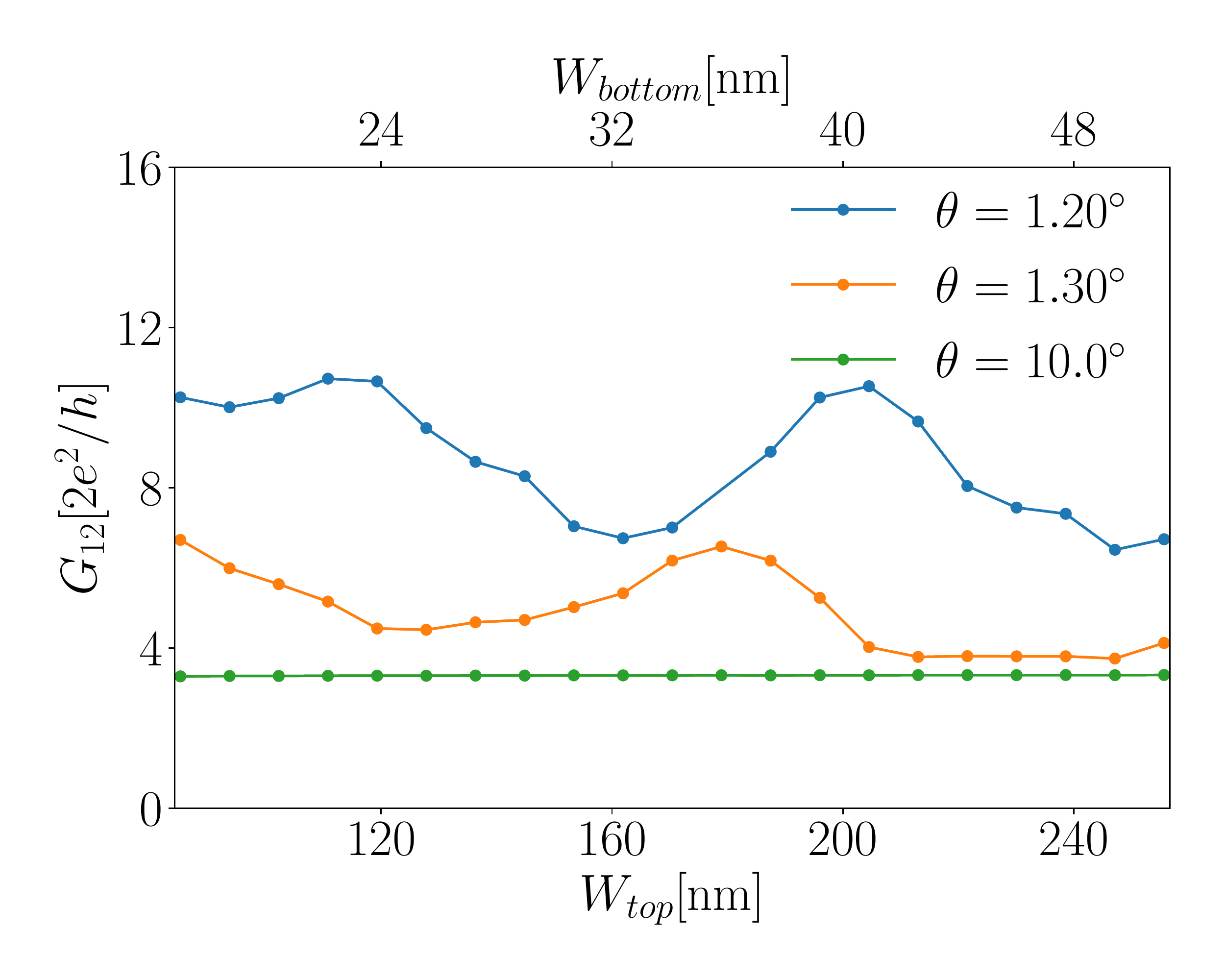}
\caption{Minimal conductance as a function of the system size: we show the wide-junction conductance $G_{12}$ at the Dirac-point energy $E=0$ for small and large angles. We have fixed the aspect ratio of the systems to $W_\mathrm{top}/W_\mathrm{bottom}=5$.}
\label{fig:minimal_conductance_with_size}
\end{figure}

\co{Discuss the minimal conductance}

Finally, in Fig.~\ref{fig:minimal_conductance_with_size} we compare the minimal conductances of our setup for small and large twist angles as a function of the system size with fixed aspect ratio.
For large angles, the minimal conductance converges quickly to the universal single-layer graphene value, which is in agreement with an effective decoupling of the layers.
On the contrary, as we get closer to the magic angle the minimal conductance is enhanced and develops pronounced oscillations as a function of the system size.
Notably, the minimal conductance in this regime is much larger than the universal value of Dirac-point conductance for a single layer indicating that both layers now contribute to the electronic transport.

\section{Conductance signatures of van Hove singularities}
\label{sec:conductance-and-VHS}

\begin{figure*}[t]
\includegraphics[width=\textwidth]{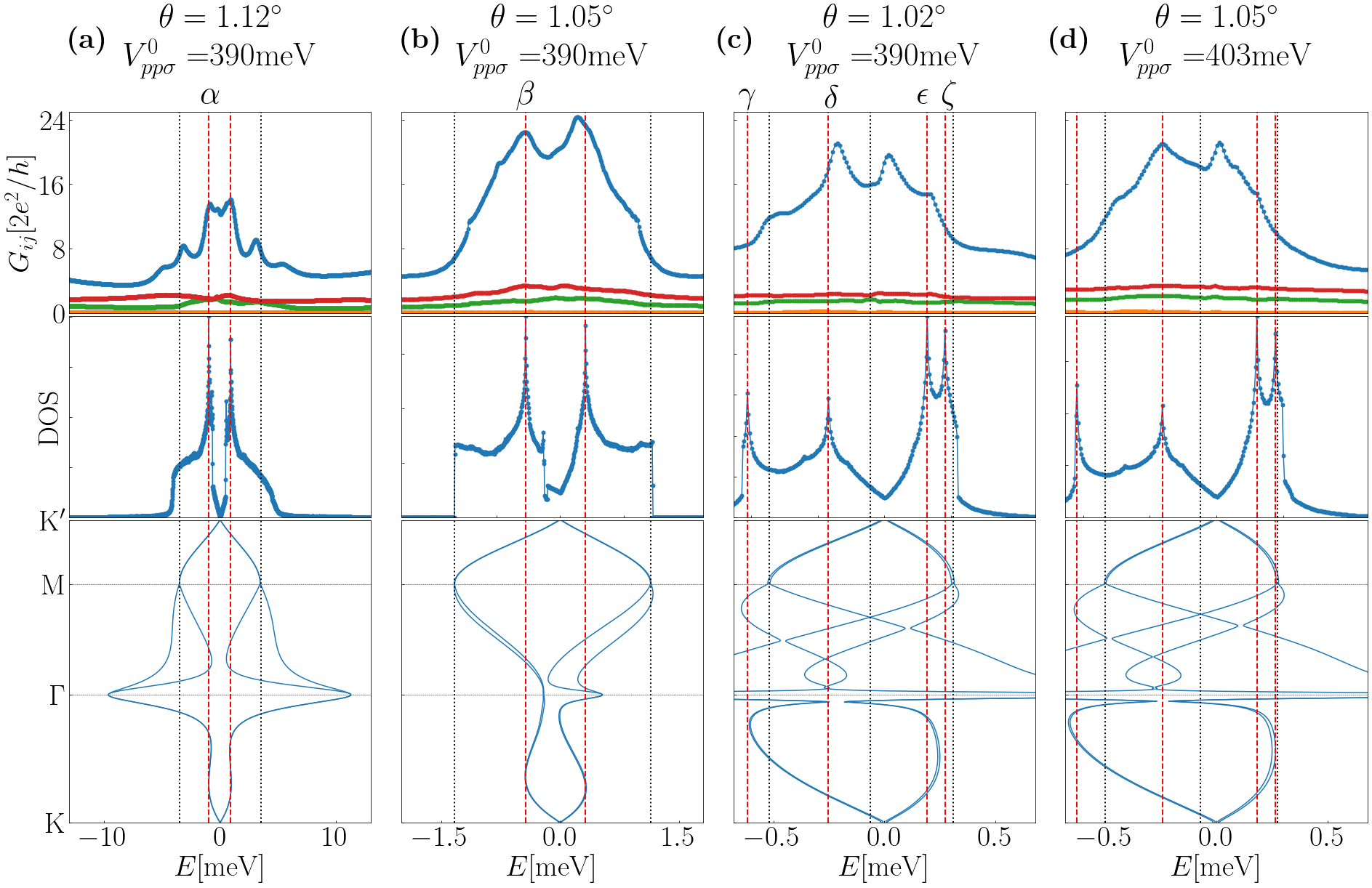}
\caption{Low-energy conductance, bulk density of states, and bulk energy bands for four selected twist angles $\theta$ and interlayer hopping amplitudes $V_{pp\sigma}^0$:
The first row shows the four independent components of the conductance tensor $G_{ij}$ as a function of energy $E$: $G_{12}$ (blue), $G_{34}$ (orange), $G_{13}$ (green), and $G_{14}$ (red).
In the second row, we plot the bulk DOS.
The third row shows the bulk spectrum along along high-symmetry lines of the moir{\'e} BZ.
Red dashed lines indicate the energetic positions of selected VHSs, whereas black dotted lines indicate other, non-singular crossings of energy bands discussed in the text. The Fermi surface features of the VHSs labelled $\alpha$--$\zeta$ are illustrated in Fig.~\ref{fig:fermi_surfaces}.}
\label{fig:conductance_dos_spectrum}
\end{figure*}

\co{We now discuss conductance as a function of energy at fixed twist angles and fixed interlayer couplings}

Next, we compare the conductance $G_{ij}(E)$ of the TBLG device to the bulk spectrum of TBLG focusing on the energy regime around the quasi-flat bands.
Fig.~\ref{fig:conductance_dos_spectrum} shows our results for a few selected cases.
To better link the features in the conductance to the DOS and the dispersion of the energy bands, we show the evolution of the Fermi surface at energies around these features in Fig.~\ref{fig:fermi_surfaces}. More results can be found in the Supplemental Material (SM)~\cite{supp}.
We note that with decreasing twist angle, additional energy band crossings lead to a plethora of features in tiny regions of the moir{\'e} BZ.
Even though we see some of these features as peaks in our finite-momentum grid calculations of the DOS, we do not observe conductance signatures that can be attributed to them for the system sizes considered. 
In the following, we will therefore restrict the discussion regarding van-Hove singularities to the most pronounced peaks in the computed DOS.
We have checked that all of the discussed peaks correspond to ordinary van Hove singularities with a logarithmic divergence.

\subsection{Twist angles $\theta \geq 1.12^\circ$}
\label{sec:conductance-and-VHS_A}

\co{Discuss the angle regime that leads up to the first conductance maximum in Fig.~2(a). Discuss the main features that can be linked to double VHSs.}

\begin{figure*}[t]
\includegraphics[width=\textwidth]{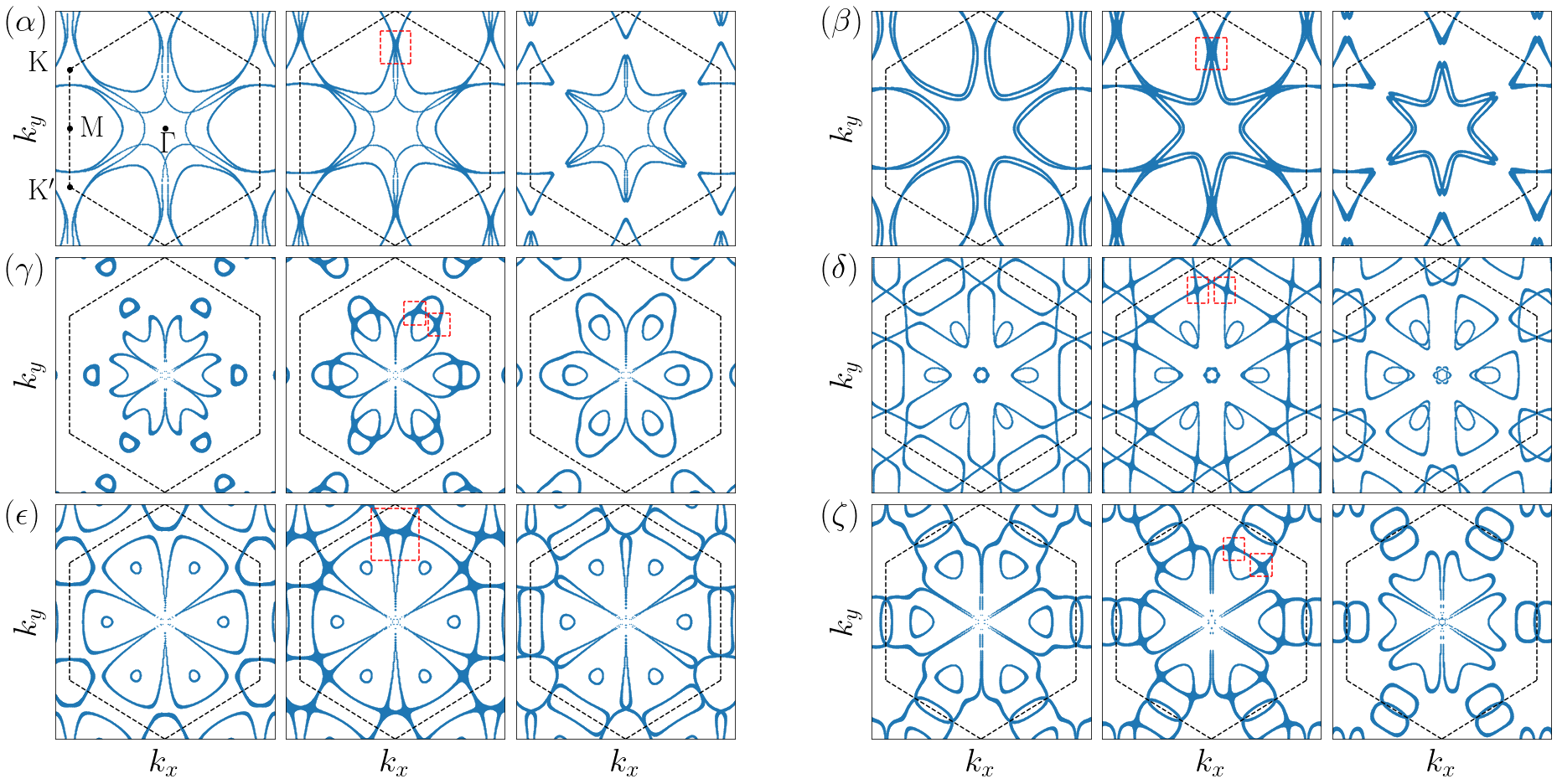}
\caption{Evolution of Fermi surfaces around the VHSs labelled ($\alpha$) to ($\zeta$) in Fig.~\ref{fig:conductance_dos_spectrum}:
For each case, we show the Fermi surfaces at energies slightly below the VHS, at the VHS, and slightly above the VHS (from left to right).
The dashed black line indicates the corresponding moir{\'e} Brillouin zone.
The red boxes highlight band crossings corresponding to VHSs.
In the first panel of ($\alpha)$, we further show the high-symmetry points used in Fig.~\ref{fig:conductance_dos_spectrum}.}
\label{fig:fermi_surfaces}
\end{figure*}

We begin by discussing the conductance in the small-angle regime leading up to the first pronounced maximum in the minimal conductance $G_{12}(E=0)$ at $\theta=1.14^\circ$ [see Fig.~\ref{fig:conductance_angle_and_interlayer_dependance}(a)].
Fig.~\ref{fig:conductance_dos_spectrum}(a) shows the conductance, the DOS, and the energy bands at the closest commensurate angle $\theta=1.12^\circ$.
As we decrease the twist angle to small values, the energy bands of the system are reconstructed due to the enlargement of the moir{\'e} unit cell.
This reconstruction flattens the Dirac cones at the $K$ and $K'$ points of the moir{\'e} BZ and eventually decouples the four corresponding spin-degenerate bands from the rest of the bulk bands thereby forming isolated quasi-flat bands.
Close to the Dirac point energy $E=0$, we observe two pronounced peaks in the wide-junction conductance $G_{12}$ that align with the two main peaks of the bulk DOS.
These peaks originate from VHSs in the bulk energy bands.
As illustrated in Fig.~\ref{fig:fermi_surfaces} [see panels ($\alpha$)] for one of the singularities, each of them corresponds to a \emph{double saddle point} involving two energy bands crossing along the $\Gamma K$ lines of the BZ.
At these points, the topology of the Fermi surface changes constituting a Lifshitz transition leading to a measurable transport signature.
Similar signatures arising from Lifshitz transitions associated with van Hove singularities have been experimentally probed in the context of untwisted bilayer graphene~\cite{Davydov2019, Suszalski2019, Jayaraman2021}.

In agreement with the literature, the DOS goes to zero at the Dirac point energy, which is similar to monolayer graphene.
In monolayer graphene the conductance still takes a finite value, as discussed above.
Here, however, the conductance $G_{12}$ at $E=0$ is large and exceeds the monolayer and the Bernal-stacked bilayer value~\cite{note_bilayer_conductance} considerably, as we already pointed out above.
This is mainly due to the strong coupling of the layers in this regime, so the Dirac points of both layers now contribute to the conductance. Another reason for the enhancement is the broadening of the DOS peaks in the conductance of the device: as the bands become flatter, the two broadened VHS features move closer together and their overlap at $E=0$ is enhanced.
These are factors contributing to the increase of $G_{12}(E=0)$ towards smaller angles seen in Fig.~\ref{fig:conductance_angle_and_interlayer_dependance}(a).

\co{Discuss the smaller satellite peaks in this regime.}

We further observe two smaller peaks around the two main peaks in the wide-junction conductance $G_{12}$.
In contrast to the main peaks, these do not correspond to distinct features in the DOS and, therefore, do not correspond to VHSs.
Nevertheless, we find that they instead correspond to a band crossing at the $M$ points in the BZ (see Appendix~\ref{app:Fermi_surfaces_non-singular_crossings}).
These band crossings constitute Lifshitz transitions that are not associated with VHSs.

\subsection{Twist angles $1.12^\circ > \theta \geq 1.05^\circ$}

\co{Discuss the regime between the two maxima in the zero-energy conductance down to the magic angle.}

As we lower the twist angle further from $\theta=1.12^\circ$ down to the first magic angle at $\theta=1.05^\circ$, the bandwidth of the quasi-flat bands is reduced by nearly one order of magnitude [see Fig.~\ref{fig:conductance_dos_spectrum}(b)].
This enhances the DOS overall and, consequently, leads to a generally larger wide-junction conductance $G_{12}$ within the energy range of the isolated quasi-flat bands.
We still observe two pronounced main peaks close to $E=0$, which correspond to the same type of double-saddle point VHSs as before [see panels ($\beta$) in Fig.~\ref{fig:fermi_surfaces}].
The band crossings responsible for the additional conductance features at $\theta=1.12^\circ$ have moved to the band edges of the isolated quasi-flat bands and no longer stand out as much as before.
We merely see small kinks in the $G_{12}$ conductance close in energy to these features.
At the Dirac point energy, the DOS is no longer zero because of other parts of the moir\'{e} bands crossing this energy.
As before, the wide-junction conductance $G_{12}$ is largely enhanced at $E=0$ and almost one order of magnitude larger than the monolayer and Bernal-stacked bilayer value.
This is due to several factors: the strong coupling between the layers, the broadening of the VHS features, and also additional states crossing the Dirac-point energy.
Besides the two main peaks in the DOS, we note that there are also two smaller features in the DOS slightly below the Dirac point energy.
We find that these correspond to ordinary VHS originating from saddle points close to the center of the BZ (see SM~\cite{supp}).  
They are not clearly visible in the wide-junction conductance $G_{12}$, because their contributions merge with the broadened features associated with the other VHSs.
Note that these are the same singularities that are visible in the DOS of Fig.~\ref{fig:conductance_dos_spectrum}(a) in between and close to the two main peaks discussed in Sec.~\ref{sec:conductance-and-VHS_A}.  

\subsection{Twist angles $1.05^\circ > \theta \geq 1.02^\circ$}

\co{Discuss angles below the magic angle}

Tuning the twist angle to the next commensurate angle $\theta=1.02^\circ$ below the magic angle [see Fig.~\ref{fig:conductance_dos_spectrum}(c)], the quasi-flat bands reconnect with the other bulk bands close to $\Gamma$.
The DOS features four pronounced peaks instead of two.
A closer analysis reveals that they correspond to different VHSs, as illustrated in the panels ($\gamma$) to ($\zeta$) in Fig.~\ref{fig:fermi_surfaces}, connected to different local transitions of the Fermi surface topology.
However, we find that only the VHS indicated by ($\delta$) in Fig.~\ref{fig:fermi_surfaces} close to the Dirac-point energy leads to a clear feature in the wide-junction conductance $G_{12}$.
The two VHSs indicated by ($\epsilon$) and ($\zeta$) above the Dirac points are too close in energy to be resolved separately and, together, they only cause a small kink in $G_{12}$.
We note that the band crossing at $M$, leading to a small conductance feature at larger angles [see Fig.~\ref{fig:conductance_dos_spectrum}(a)], is close in energy to these VHSs and, therefore, might also contribute to the kink in $G_{12}$.
Similarly, the related band crossing at $M$ below the Dirac point is close in energy to the VHS at ($\gamma$) and, therefore, we attribute them together to a small bump in the $G_{12}$ conductance within this energy range.

\co{Discuss the feature that doesn't seem to correspond to anything}

In addition, we find another pronounced peak in the $G_{12}$ conductance roughly at the Dirac-point energy.
This feature is relatively far in energy from all VHSs and, therefore, does not correspond to a singularity.
Nevertheless, we note that it is close in energy to a tangential band touching point along the $\Gamma M$ high-symmetry lines of the moir{\'e} BZ (see Appendix~\ref{app:Fermi_surfaces_non-singular_crossings}).

Generally, we observe discrepancies between the energetic positions of features in the conductance and the associated spectral features of up to $0.1\,\mathrm{meV}$ for the TBLG devices considered.
For angles $\theta\geq 1.05^\circ$, such discrepancies are small with respect to the width of the quasi-flat bands.
For smaller angles, the features of the quasi-flat bands move even closer together leading to a larger relative discrepancy in this regime. 
In Appendix~\ref{app:size_effects}, we analyze how the system size affects the broadening and the energetic positions of the conductance features.  

\subsection{Evolution of van Hove singularities between commensurate angles}

\co{Briefly discuss the last column of Fig.3, where we vary the interlayer coupling instead.}
Instead of lowering the twist angle, we can alternatively increase the nearest-neighbor interlayer coupling $V_{pp\sigma}^0$ to approximate the bulk spectrum at the next commensurate twist angle.
We demonstrate this in Fig.~\ref{fig:conductance_dos_spectrum}(d) showing that the bulk energy bands and the bulk DOS are almost identical to those in Fig.~\ref{fig:conductance_dos_spectrum}(c).
Also the conductances are in good qualitative agreement despite minor quantitative deviations.
This makes it possible to unravel the evolution of VHSs between two commensurate twist angles.

\co{Briefly describe the evolution of the VHSs between the two twist angles.}

We have used the interlayer hopping to continuously tune between the energy bands and DOS spectra corresponding to the commensurate angles $\theta=1.05^\circ$ and $1.02^\circ$ (see SM~\cite{supp}).
Starting from the magic angle $\theta=1.05^\circ$ [see Fig.~\ref{fig:conductance_dos_spectrum}(b)], the two VHSs below the Dirac point energy close to $\Gamma$ merge, which is accompanied by changes in the Fermi surface topology close to the $\Gamma$ point.
We observe that a single VHS emerges from the two original VHSs, which moves further up in energy.
It passes through the Dirac-point energy and eventually merges with the large double-saddle point VHS feature.
The energy bands associated with the two singularities reconnect and form new singularities, which subsequently split in energy.
These are the two pronounced VHS features we observe at $\theta=1.02^\circ$ for positive energies.
On the other hand, the double-saddle point VHS below the Dirac-point energy we observe at the magic angle $\theta=1.05^\circ$ splits into two separate VHSs in the DOS.

\co{Some general remarks about using this procedure for other angles.}

A similar analysis can be performed to study the evolution of energy bands and VHSs between any two commensurate twist angles provided that the angle difference is sufficiently small.
In particular, this applies to the small angle regime below the magic angle where the distance between adjacent commensurate angles rapidly goes to zero.

\section{Nonlinear transport and dynamics \label{device-applications}}

The predicted nonlinear current response to an applied voltage, following from the energy dependence of the conductance, suggests that TBLG systems can be utilized in various kinds of generators, frequency multipliers, frequency mixers, parametric amplifiers and detectors of electromagnetic radiation~\cite{Tien-Gordon,Tucker79, Tucker85, WACKER02, Renk05, Hayton13, Kirill22}.
Here, we briefly comment on some aspects regarding potential applications of TBLG devices in the light of our transport results.

The simplest approach to describe the photon-assisted current $I(t)$ as a response to a time-dependent voltage
\begin{equation}
V(t)=V_{dc} +\sum_{k=1}^{N} V_{\omega_{k}}\cos(\omega_{k}t +
\alpha_{k}) \label{Voltage-t}
\end{equation}
is based on the formula
\begin{eqnarray}
I(t)&=&\sum_{n_1,...,n_N}
\sum_{m_1,...,m_N} \bigg[ \prod_{k=1}^N J_{n_k} (
\beta_k) J_{n_k+m_k}( \beta_k) \bigg] \nonumber\\ && \hspace{-0.5cm} \times \bigg\{
I_S \big(eV_{dc}+\sum_{k=1}^N n_k \hbar \omega_k \big)
\cos{\big[\sum_{k=1}^N m_k(\omega_k t+\alpha_k)\big]}\nonumber
\\&& \hspace{-0.5cm}+K_S \big(eV_{dc}+\sum_{k=1}^{N} n_k \hbar \omega_k \big)
\sin{\big[\sum_{k=1}^N m_k(\omega_k t+\alpha_k)\big]} \bigg\}, \nonumber \\
\label{Tucker-formulas}
\end{eqnarray}
where $J_n(x)$ are Bessel functions, the summations are from $-\infty$ to $\infty$ and $\beta_k=eV_{\omega_k}/\hbar \omega_k$. Here, $I_S(e V_{dc})$ is the static current-voltage characteristic and $K_S$ is related to it by the Kramers-Kronig relation
\begin{equation}
K_S(E)=\frac{1}{\pi} {\cal P} \int_{\infty}^\infty dE' \frac{I_S(E')}{E'-E},
\end{equation}
where ${\cal P}$ denotes the Cauchy principal value. 
The formula in Eq.~(\ref{Tucker-formulas}) is a  generalization of the Tien-Gordon-Tucker relations~\cite{Tien-Gordon, Tucker79, Tucker85} to arbitrary polychromatic fields, and it has been utilized in the description of photon-assisted transport in various quantum systems operating in sequential tunneling and miniband transport regimes as well as in Josephson junctions and exciton condensates  \cite{Tucker79, Tucker85, WACKER02, PLATERO04, Koval04, Hyart13}. 
From Eq.~(\ref{Tucker-formulas}) it is easy to notice that for the type of applications discussed above it would be preferable to have large conductance and strongly nonlinear $I_S(eV_{dc})$ characteristics. Namely, the magnitudes of the Fourier components of $I(t)$ depend, in addition to the amplitudes of the driving fields, on the overall conductance.
Additionally, it follows from the properties of the Bessel functions that for linear $I_S(eV_{dc})$ characteristics the current response would be $I(t)=G_0 V(t)$, where $G_0$ is the conductance. Thus, linear $I_S(eV_{dc})$ characteristics are unsuitable for the use in generators, frequency multipliers, frequency mixers, parametric amplifiers or the type of detectors discussed in Refs.~\onlinecite{Tucker79, Tucker85, WACKER02, Renk05, Hayton13, Kirill22}. We also point out that for some applications it is advantageous if the energy scale of nonlinearities in the $I_S(eV_{dc})$ characteristics roughly matches the photon energies $\hbar \omega_k$. According to our calculations, 
TBLG devices close to the magic angle typically have almost an order of magnitude larger conductance than monolayer graphene systems of similar size. Moreover, our calculations demonstrate that there exists a strong quantum nonlinearity due to the van Hove singularities. Furthermore, we point out that the energy scales of the nonlinearities vary within the range $0.1-1\,\mathrm{meV}$, so that TBLG devices are promising also for high-frequency applications in the range of frequencies from $100\,\mathrm{GHz}$ to $1\,\mathrm{THz}$. This frequency range is particularly important from the technological perspective because of the lack of compact solid-state technologies for generating and detecting THz  radiation.

\section{Conclusion}
\label{sec:conclusion}

\co{Summarize results.}

We have studied low-energy, four-terminal conductance of twisted bilayer graphene in mesoscopic samples formed by crossing a narrow and a wide graphene nanoribbon.
At large twist angles, the interlayer conductance vanishes because the two junctions effectively decouple.
At small twist angles, however, the interlayer conductance becomes nonzero indicating coupling between the graphene sheets, whereas the narrow-junction conductance goes to zero.
Notably, the wide-junction conductance is largely enhanced due to the formation of quasi-flat energy bands. We have shown that the conductance can also be controlled with pressure.

We find that the low-energy quantum transport close to the magic angle is affected by several factors.
We have identified VHSs within the quasi-flat bands contributing to conductance features through their effect on the bulk DOS. 
We observe that such features are generally pronounced around the center of the quasi-flat bands, whereas they are suppressed closer to their edges.
Moreover, we find additional peaks in the conductance not corresponding to features in the bulk DOS.
These peaks coincide in energy with non-singular band crossings located at high-symmetry points and lines of the bulk BZ.
Finally, we also observe a broadening of the various conductance features.
As a consequence of the broadening and of the strong coupling between the layers, the quasi-flat bands exhibit a large minimal conductance at the Dirac-point energy exceeding the value of single-layer and Bernal-stacked bilayer graphene considerably.
We have further studied the continuous evolution of VHSs corresponding to different commensurate twist angles by tuning the interlayer coupling, which could experimentally be realized by applying pressure.

\co{Discuss significance of our results and possibly give an outlook.}

VHSs play a prominent role in the exotic phenomena of magic-angle twisted bilayer graphene by amplifying electronic correlation effects. 
Our results indicate that the VHSs strongly influence also the transport properties of this system, but we find that there are additional factors affecting them.
Moreover, our calculations provide an estimate on the sample sizes required to resolve signatures of the flat-band VHSs in the conductance.
Specifically, we have considered bilayer samples of size $50\,\mathrm{nm}\times 250\,\mathrm{nm}$.
Even though our study neglects interaction effects, our results are also directly relevant experimentally:  
in a transport-measurement setup the appearance of correlated phases could be avoided, for instance, by tuning the Fermi level away from the quasi-flat energy bands and by using voltages to probe energies within the flat bands, or by operating at temperatures above the critical temperatures of the correlated phases.
In the case of extremely accurate low-temperature transport experiments the interactions could also be screened~\cite{Veyrat2020, Stepanov2020}. We have neglected the effects of disorder as they are expected to be unimportant for the system sizes considered.
We have further disregarded lattice relaxation effects, which may play a role for twist angles $\lesssim 1^\circ$ due to the possible formation of alternating AB and AA stacking domains instead of the moir\'e pattern~\cite{Nam2017}.

As an interesting future research direction, we propose the investigation of TBLG devices for the use as compact solid-state frequency multipliers, frequency mixers, parametric amplifiers and detectors operating at THz frequencies. The high sensitivity of the conductance to external parameters suggests that TBLG devices could be utilized also in various types of sensitive detectors.

\begin{acknowledgments}
\textit{Acknowledgments:}
The research was partially supported by the Foundation for Polish Science through the IRA Programme co-financed by EU within SG OP. T. H.  acknowledges
the computational resources provided by
the Aalto Science-IT project and the 
financial support from the 
Academy of Finland Project No.
331094.
A. L. acknowledges support from a Marie Sk{\l}odowska-Curie Individual Fellowship under grant MagTopCSL (ID 101029345).
J.T. received founding from the National Science Centre, Poland, within the QuantERA II Programme that has received funding from the European Union’s Horizon 2020 research and innovation programme under Grant Agreement No 101017733, Project Registration Number: 2021/03/Y/ST3/00191, acronym TOBITS. We acknowledge the access to the computing facilities of the Interdisciplinary Center of Modeling at the University of Warsaw, Grant No. G86-1064.

\end{acknowledgments}

\textit{Data availability:} The data shown in the figures is available at Ref.~\onlinecite{zenodo}.

\bibliography{bibliography}

\begin{thebibliography}{73}
\bibitem [{not()}]{note_bilayer_conductance}%
  \BibitemOpen
  \href@noop {} {}\bibinfo {note} {We note that the universal value of the
  two-terminal minimal conductivity of Bernal-stacked bilayer graphene is
  identical to the monolayer value if single-layer leads are used that are
  connected to only one of the layers of the scattering region.}\BibitemShut
  {Stop}%
\end{thebibliography}

\appendix

\renewcommand{\thefigure}{S\arabic{figure}}
\setcounter{figure}{0}    

\renewcommand{\theequation}{S\arabic{equation}}
\setcounter{equation}{0}  

\section{Non-singular band-crossing features in the conductance}
\label{app:Fermi_surfaces_non-singular_crossings}

\begin{figure}[h]
\includegraphics[width=0.48\textwidth]{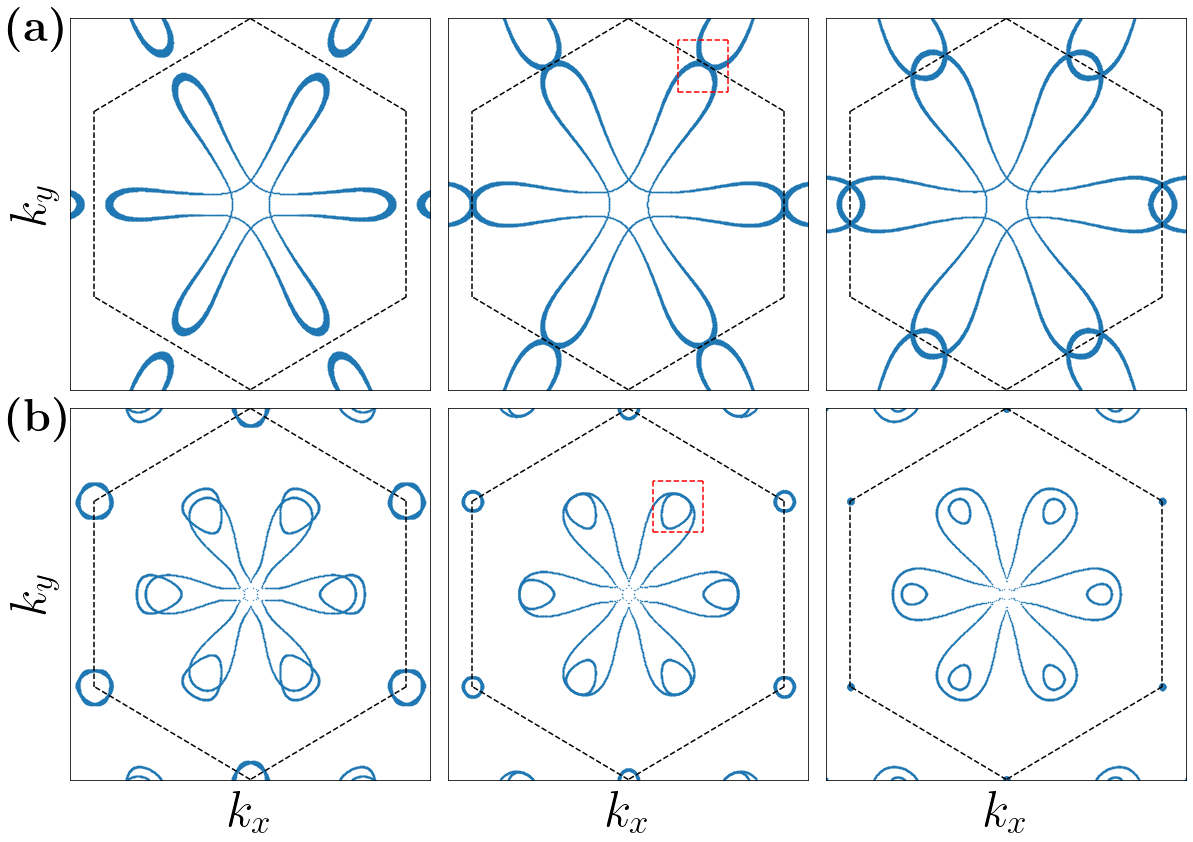}
\caption{Evolution of Fermi surfaces around band crossings in the Brillouin zone: (a) band crossing at M, (b) band crossing along $\Gamma$M.
For each case, we show the Fermi surfaces at energies slightly below the crossing, at the crossing, and slightly above the crossing (from left to right).
The dashed red line indicates the corresponding moir{\'e} Brillouin zone.}
\label{fig:fermi_surfaces_appendix}
\end{figure}

\co{Discuss very briefly the additional features we see in the conductance and that they coincide in energy with band crossing points at M or along $\Gamma$M.}

In Sec.~\ref{sec:conductance-and-VHS}, we identified non-singular band crossing points that approximately coincide in energy with peaks in the wide junction conductance.
Here, we present the evolution of the Fermi surfaces for some of these crossing points.

Figure~\ref{fig:fermi_surfaces_appendix}(a) shows a band-crossing transition at the $M$ point in the BZ.
The corresponding energy of the crossing point is highlighted by the dashed black line below $E=0$ in Fig.~\ref{fig:conductance_dos_spectrum}(a) of the main text.
With increasing energy, the Fermi surface extends and intersects with itself at the $M$ point.
A similar band-crossing transition takes place above $E=0$ where the Fermi surface intersects with itself at the $M$ point (not shown).
We find the same band-crossing transitions at the $M$ point also in Figs.~\ref{fig:conductance_dos_spectrum}(b-d).

In Figs.~\ref{fig:conductance_dos_spectrum}(c) and~(d), we further identified a feature in the wide junction conductance close to $E=0$, which is far in energy from any VHS.
Instead, it is close in energy to a band crossing point, whose energetic position is highlighted by the dashed black line in Fig.~\ref{fig:conductance_dos_spectrum}(c) close to $E=0$.
The corresponding evolution of the Fermi surface is illustrated in Fig.~\ref{fig:fermi_surfaces_appendix}(b).
With increasing energy, the circular feature along $\Gamma M$ close to the center of the BZ grows, touches another Fermi line tangentially, and then crosses through it.

\section{Effects of system size on the conductance}
\label{app:size_effects}

\begin{figure}[b]
\includegraphics[width=0.4\textwidth]{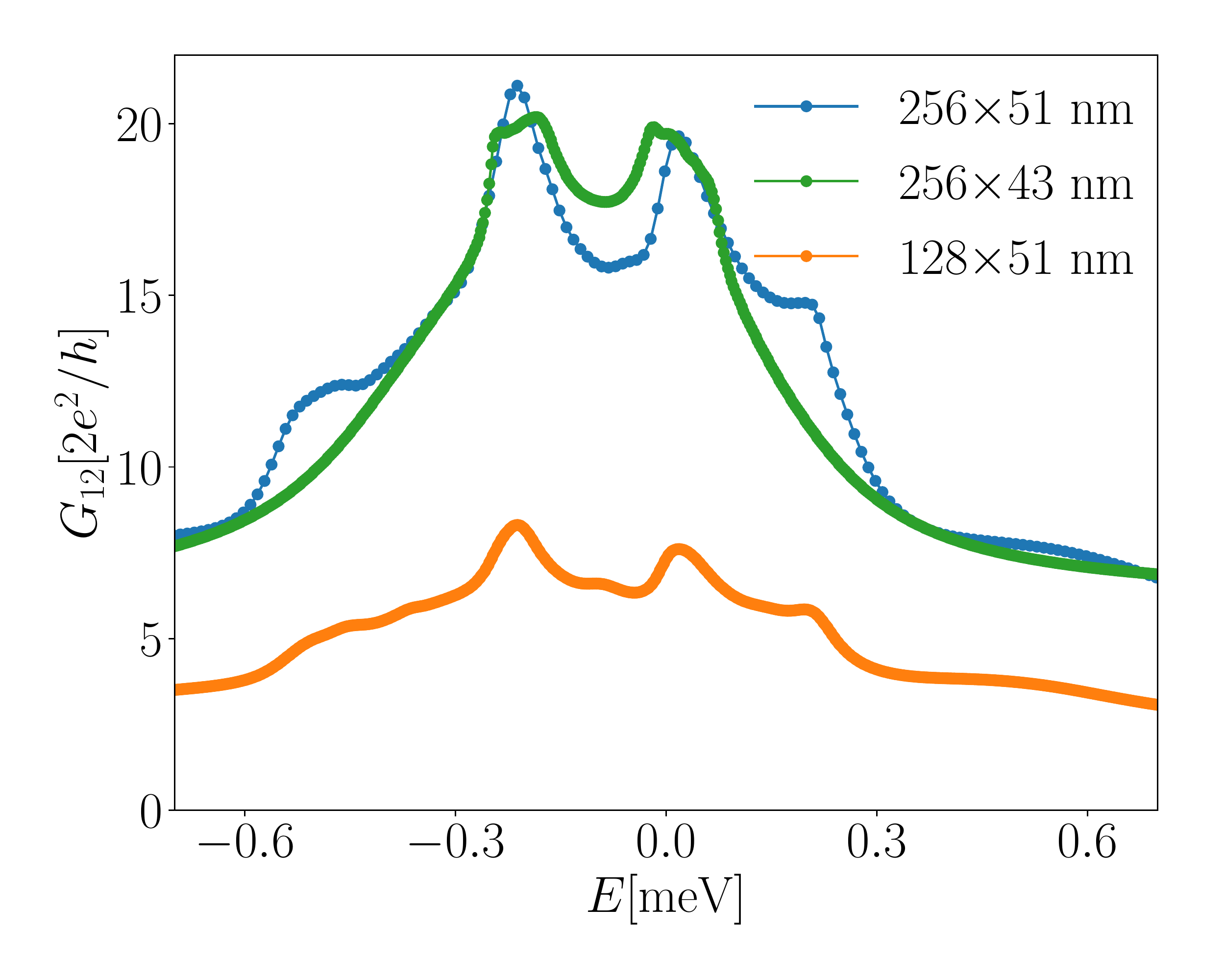}
\caption{Effects of system size on conductance: we show the wide junction conductance $G_{12}$ as a function of energy $E$ for different device dimensions $W_\mathrm{top} \times W_\mathrm{bottom}$.}
\label{fig:finite-size_effects_appendix}
\end{figure}

In this section, we analyze how a change of the device dimensions affects the conductance.
In particular, we independently vary the width $W_\mathrm{bottom}$ of the long bottom junction and the width $W_\mathrm{top}$ of the wide top junction of our TBLG device (see Fig.~\ref{fig:setup}).
We focus on the wide junction conductance $G_{12}$ and discuss, as an example, a device with a twist angle of $\theta=1.02^\circ$ and interlayer hopping $V_{pp\sigma}^0=390\,\mathrm{meV}$ [compare to Fig.~\ref{fig:conductance_dos_spectrum}(c)].

We show our results in Fig.~\ref{fig:finite-size_effects_appendix}.
We find that decreasing the wide junction width $W_\mathrm{top}$ suppresses the conductance overall, which is due to the reduced DOS in the top leads.
All the conductance features are still visible.
Nevertheless, there is no effect on the energetic position and width of the conductance features.
On the contrary, a change of the long-junction width $W_\mathrm{bottom}$ affects the broadening and also the energetic position of the peaks.
As $W_\mathrm{bottom}$ gets smaller, the features become flatter and more smeared out.
In particular, we observe that the kink-like features close to the edges of the flat band disappear for a slightly narrower system.

\end{document}